\documentclass[preprint]{aastex}
\begin{document}

\newcommand{\gsim}{\mbox{\raisebox{-1.0ex}{$~\stackrel{\textstyle >}
{\textstyle \sim}~$ }}}
\newcommand{\lsim}{\mbox{\raisebox{-1.0ex}{$~\stackrel{\textstyle <}
{\textstyle \sim}~$ }}}
\newcommand{\psim}{\mbox{\raisebox{-1.0ex}{$~\stackrel{\textstyle \propto}
{\textstyle \sim}~$ }}}
\newcommand{\vect}[1]{\mbox{\boldmath${#1}$}}
\newcommand{\lmk}{\left(}
\newcommand{\rmk}{\right)}
\newcommand{\lnk}{\left\{ }
\newcommand{\nn}{\nonumber}
\newcommand{\rnk}{\right\} }
\newcommand{\lkk}{\left[}
\newcommand{\rkk}{\right]}
\newcommand{\lla}{\left\langle}
\newcommand{\p}{\partial}
\newcommand{\rra}{\right\rangle}
\newcommand{\vex}{{\vect x}}
\newcommand{\vek}{{\vect k}}
\newcommand{\vel}{{\vect l}}
\newcommand{\vem}{{\vect m}}
\newcommand{\ven}{{\vect n}}
\newcommand{\vep}{{\vect p}}
\newcommand{\veq}{{\vect q}}
\newcommand{\veX}{{\vect X}}
\newcommand{\veV}{{\vect V}}
\newcommand{\beq}{\begin{equation}}
\newcommand{\eeq}{\end{equation}}
\newcommand{\beqa}{\begin{eqnarray}}
\newcommand{\eeqa}{\end{eqnarray}}
\newcommand{\mpc}{\rm Mpc}
\newcommand{\hmpc}{{h^{-1}\rm Mpc}}
\newcommand{\ch}{{\cal H}}
\newcommand{\lab}{\label}
\newcommand{\calv}{{\cal V}}

\title{Numerical Analyses of Weakly Nonlinear Velocity-Density Coupling}

\author{\sc Naoki Seto}
\affil{Department of Earth and Space Science,
Osaka University,
Toyonaka 560-0043, Japan
}

\author{\sc Naoshi Sugiyama}
\affil{Division of Theoretical Astrophysics,
National Astronomical Observatory, 181-8588, Japan
}

\begin{abstract}
We study evolution of various statistical
 quantities  of smoothed  cosmic  density and velocity fields
 using N-body simulations.  The parameter
 $C\equiv \lla\veV^2\delta\rra/(\lla\veV^2\rra \lla\delta^2\rra)$
 characterizes   nonlinear coupling of these two
 fields and determines  behavior of bulk velocity dispersion as a
 function of local density contrast. 
 It is found that this parameter depends strongly on the
smoothing scale  even in quasi-linear regimes where
 the skewness parameter
 $S_3\equiv \lla\delta^3\rra/\lla\delta^2\rra^2 $  is
 nearly  constant and close to  the predicted value  by
 the second-order perturbation theory.  We also analyze weakly nonlinear
 effects caused by an  adaptive smoothing  known as the gather
 approach. 
\end{abstract}

\section{INTRODUCTION}

It is now commonly accepted that the large-scale structure in the
universe is gravitationally evolved from small initial
inhomogeneities. In addition to redshift surveys of galaxies and
temperature anisotropies of the cosmic microwave background radiation,
observational analysis of the cosmic velocity field would bring us
fruitful information of our universe, such as, the density parameter
$\Omega_0$ or the matter
 power spectrum (Dekel 1994, Strauss \& Willick 1995).
In 
order to analyze 
the velocity field, however, we need to clarify a basic problem, whether
the velocity field of galaxies is statistically different from that of
dark-matter particles.  The observed line-of-sight
peculiar velocity field is not usually traced by the underlying density
field but by galaxies, since the measurements of the distances are
crucial element to determine the peculiar velocity field and are usually
carried by using galaxies.  This important problem in observational
cosmology is broadly called the velocity bias.  The velocity bias is
often studied using semi-analytic models of galaxy formation (Cen \&
Ostriker 1992, Kauffmann et al. 1999, Narayanan et al. 2000). But it is
not easy to make definite predictions with this approach, because our
understanding of astrophysical process related to galaxy formation
reaches far from a quantitative level.

The number density of galaxies is expected to be closely related to
the density contrast of dark matter.   Therefore we can obtain some
insights 
about the velocity bias by analyzing relation between the velocity and
density fields only  of dark matter particles.   
This analysis is relatively 
simple,  as only  gravitation is the relevant physical process
for their evolution.   When the 
initial fluctuations are isotropic random Gaussian distributed, these
two fields 
at a same point 
 are statistically independent in the framework of 
 the linear perturbation theory.
One of the author (Seto 2000a) studied weakly nonlinear evolution of the
bulk (smoothed) velocity dispersion $\Sigma_V^2(\delta)$ as a function of
local smoothed
density contrast $\delta$.  He used the Eulerian 
 second-order perturbation theory  and
the Edgeworth expansion method, and  found that the constrained velocity
dispersion 
$\Sigma_V^2(\delta)$ 
 is written in the form   $\Sigma_V^2(\delta)\propto
(1+C\delta+O(\delta^2))$ with a parameter $C\simeq 0.2\sim 0.3$ for
typical CDM models (see also Chodorowski \& {\L}okas 1997 for the
velocity divergence field).

However, numerical results given in Kepner et
al. (1997) show that magnitude of bulk velocity is almost independent
on the local density contrast $\delta$  at nonlinear
regimes (see their figs 2.a 
and 3.a).  This behavior shows
 a remarkable difference from the above second-order prediction.
Therefore we expect that the quantities $C$ and $\Sigma^2_V(\delta)$ 
 show interesting behaviors
by  changing a spatial scale from linear to nonlinear, and would be
useful measures to check  performance of  the perturbative analysis for weakly
nonlinear evolution of  
the large-scale structure.  In
this article we study these quantities using N-body simulations and
compare numerical  results with analytical second-order 
predictions.

This article is organized as follows.
We begin by summarizing the second-order analysis of Seto (2000a) in \S
2.1.  Smoothing operation is crucial for our numerical analysis and its
proper treatment is very important to make a quantitative analysis. There
are some variations for smoothing methods.  A
mass-weighted smoothing scheme is convenient for analyzing N-body
data  and an adaptive
smoothing scheme  might be efficient to resolve cosmic
fields,  especially
in sparse underdense regions. We  perturbatively study these methods  in \S 2.2 and \S 2.3.
   In \S 3 we derive an expression to estimate 
 the sampling fluctuation expected in our numerical
analysis. Numerical results are given in \S 4. Finally \S 5 is devoted
to a summary.

\section{SECOND-ORDER PERTURBATION THEORY}
\subsection{Smoothed Velocity Dispersion as a Function of Local Density}
In this subsection we  summarize
the perturbative analysis of density and velocity fields and their 
correlations given in Seto (2000a). 
We denote the volume-weighted smoothing operator $[\cdot]_R$ with
 radius $R$ for a  field $f(\vex)$ as follows:
\beq
[f(\vex)]_R\equiv \int d^3 x'  f(\vex) W(\vex'-\vex;R) \lab{vw} , 
\eeq
where $W(\vex,R)$ is a smoothing filter function.  For simplicity we also use
a notation $f_R(\vex)\equiv [f(\vex)]_R$. In this article we only employ
the Gaussian filter defined as
\beq
W(\vex;R)=\frac1{({2\pi R^2})^{3/2}} \exp\lmk -\frac{|\vex|^2}{2R^2}
\rmk, \lab{gf}
\eeq
and  discuss   velocity and density fields
smoothed by this 
filter.

The smoothed (bulk) velocity dispersion $\Sigma_V^2(\delta)$ for points
$\vex$ with 
a given smoothed overdensity $\delta_R(\vex)=\delta$ is formally
expressed as 
\beq
\Sigma_V^2(\delta)=\frac{\lla
  \veV_R(\vex)^2\delta_{D}[\delta_R(\vex)-\delta]\rra }{\lla
  \delta_{D}[\delta_R(\vex)-\delta]\rra},
\eeq
where $\delta_{D}(\cdot)$ is the Dirac's delta function and the bracket
$\lla\cdot\rra$ represents  an ensemble average.
 Let us evaluate this expression  using the second-order Eulerian
perturbation theory (Peebles 1980) and the Edgeworth expansion method
(Matsubara 1994, Juszkiewicz et al. 1995, Bernardeau \& Kofman 1995).
 We first expand the density and velocity fields as
\beqa
\delta(\vex)&=&\delta_{1}(\vex)+\delta_2(\vex)+\delta_3(\vex)+\cdots,\\
\veV(\vex)&=&\veV_1(\vex)+\veV_{2}(\vex)+\veV_3(\vex)+\cdots,
\eeqa
where $\delta_{1}(\vex)$ and $\veV_1(\vex)$ are  the linear modes,
$\delta_2(\vex)$ and $\veV_{2}(\vex)$ are the second-order modes, and so
on. We can regard the order parameter of these  expansions  as the rms
density fluctuation
$\sigma\equiv\lla\delta_R^2\rra^{1/2}$. 
We assume that the  primordial fluctuation is isotropic random-Gaussian
distributed and 
 the linear modes $\delta_1(\vex)$ and $\veV_1(\vex)$ obey this simple
statistic.  Then their
  probability distribution function $P(\delta_1,\veV_1)$
 is determined by their
covariance matrix ({\it e.g.} Bardeen et al. 1986). Although
the linear velocity field $\veV_1$ is given in terms of the linear
density
field  $\delta_1$ as $\veV_1\propto \nabla \Delta^{-1}\delta_1$, 
 this matrix is
orthogonal due to the statistical isotropy of fluctuations. Thus these
four variables (density and three components of  
velocity fields) are statistically
 independent,  as long as we discuss the density and velocity fields at a
same point.
Therefore at the linear order, velocity dispersion does not depend on
the density contrast
$\delta$ and is  given as
\beq
\Sigma_V^2(\delta)=\lla \veV_{1R}\cdot\veV_{1R}\rra+O(\sigma^4).
\eeq

Next we evaluate the higher order correction  using the multivariable Edgeworth
 expansion for the one point probability distribution
function $P(\delta,\veV)$ around its linear (gaussian) distribution.
After some tedious algebra, we obtain the following  expression up to the first
non-Gaussian correction
\beq
\Sigma_V^2(\delta)=\lla \veV_R^2  \rra(1+C\delta+O(\sigma^2)),
\lab{convel}
\eeq
where the  coefficient $C$ is defined by
\beq
C\equiv \frac{\lla \veV_R^2 \delta_R  \rra }{\lla \veV_R^2\rra \lla \delta_R^2\rra}.
\eeq 
This coefficient $C$ characterizes nonlinear couplings of the  velocity
 and the density fields. 
Equation (\ref{convel}) shows that the first nonlinear
correction for the velocity dispersion for points
 with a  given density contrast
$\delta$   is simply proportional to $C\delta$.

The leading order contributions for two factors $\lla
\veV_R\cdot\veV_R\rra$
and $\lla\delta_R^2\rra$ in the 
 denominator of $C$ are given in terms of the matter power spectrum $P(k)$ as
follows 
\beqa
\lla\delta_R^2\rra&=&\int\frac{k^2dk}{2\pi^2} P(k) W(kR)^2+O(\sigma^4),\lab{d2}\\
\sigma_V^2\equiv\lla \veV_R\cdot\veV_R\rra &=&H^2f^2\int\frac{dk}{2\pi^2} P(k) W(kR)^2+O(\sigma^4)\lab{v2},
\eeqa 
where $H$ is the  Hubble parameter and 
 $f$ is a function of cosmological parameters $\Omega_0$ and
$\lambda_0$, and  well fitted by
\beq
f\simeq \Omega_0^{0.6}+\frac{ \lambda_0}{70} \lmk 1-\frac{\Omega_0}{2}\rmk,
\eeq
in the ranges  $0.03\le  \Omega_0 \le 2 $ and $-5 \le  \lambda_0 \le 5 $
(Lahav et al. 1991). 
The function $W(kR)$ is a Fourier transformed filter
 with smoothing radius $R$. For the Gaussian 
filter (eq.[\ref{gf}]) we have $W(kR)=\exp(-k^2R^2/2)$.

The  ensemble average of an  odd function of
Gaussian variables (with vanishing means) leads to zero. We employ
the second-order perturbation theory to evaluate the first nonvanishing
contribution of the  numerator 
$\lla\veV_R\cdot\veV_R\delta_R\rra$.  
The leading order contribution for the numerator of $C$ is written as
\beq
\lla \veV_R\cdot\veV_R\delta_R\rra= \lla \veV_{1R}\cdot\veV_{1R}\delta_{2R}\rra+2\lla
\veV_{1R}\cdot\veV_{2R}\delta_{1R}\rra+O(\sigma^6).
\eeq
Using formulas for the second-order modes $\delta_2(\vex)$ and
  $\veV_2(\vex)$ ({\it e.g.}  Fry 1984, Goroff et al. 1986),
the above expression is written with the matter power spectrum
$P(k)$ as follows:
\beqa
\lla \veV_R\cdot\veV_R
\delta_R\rra&=&2H^2f^2\int_{-1}^1du\int\frac{k^2l^2dkdl}{8\pi^4}
P(k)P(l)\exp[-(k^2+l^2+klu)R^2] \nn\\
& &\times \Bigg[-\frac{u}{kl}\lnk \frac57+\frac{u}2\lmk\frac{k}{l}+\frac{l}k\rmk+\frac27
u^2 \rnk\nn\\
& &~~~ +2\frac{k+lu}{k(k^2+l^2+2klu)}\lnk \frac37+\frac{u}2\lmk\frac{k}{l}+\frac{l}k\rmk+\frac47
u^2 \rnk   \Bigg]+O(\sigma^6).\lab{vd} 
\eeqa
Here we have neglected  extremely weak dependence on the cosmological
parameters $\Omega_0$ and $\lambda_0$ that appears in the kernels of the
second-order modes.
We should notice
 that the parameter $C=O(1)$ does not depend on  normalization of the power
spectrum at its first nonvanishing contribution (see eqs.[\ref{d2}][\ref{v2}] and [\ref{vd}]).
The factors  $(Hf)^2$ cancel out between the velocity dispersion
$\sigma_V^2$ and the velocity-density moment  $\lla 
\veV_R\cdot\veV_R \delta_R\rra$. 
The parameter $C$ is affected by the
cosmological parameters
 mainly through its dependence on the power spectrum ({\it e.g.} the
shape parameter $\Gamma \simeq \Omega h$ for CDM transfer function).   

The parameter $C$ characterizes the nonlinear mode couplings
 of  velocity and
density fields. It is interesting to compare  this  with
  a same
kind of  nonlinear
quantity that is
 determined only  by the density field $\delta$. The skewness parameter
$S_3=O(1)$ is defined by the second- and third-order moments of density
contrast 
$\delta$ and written as (Peebles 1980, Fry 1984, Bouchet et al. 1992,
Juszkiewicz,  
Bouchet \& Colombi 1993, Bouchet et al. 1993, Bernardeau 1994, Baugh,
Gazta\~{n}aga \& Efstathou 1995, Kim \& Strauss 1998)
\beq
S_3\equiv \frac{\lla \delta_R^3\rra}{\lla\delta_R^2\rra^2}.
\eeq 
Here the variance $\lla \delta_R^2\rra$ is given in equation
(\ref{d2}).  For unsmoothed density field we have $S_3=34/7$ (Peebles 1980).
After some algebra we can express the third-order moment
$\lla\delta_R^3\rra$ (Gauss filter) in terms of matter power spectrum
$P(k)$ as follows 
\beqa
\lla 
\delta_R^3\rra&=6 &\int_{-1}^1du\int\frac{k^2l^2dkdl}{8\pi^4}
P(k)P(l)\exp[-(k^2+l^2+klu)R^2] \nn\\
& &\times \Bigg[\lnk \frac57+\frac{u}2\lmk\frac{k}{l}+\frac{l}k\rmk+\frac27
u^2 \rnk\nn   \Bigg]+O(\sigma^6).\lab{skew} 
\eeqa
The skewness parameter and its generalizations  are very important to
study various aspects of weakly nonlinear density field. With these
parameters  we
can discuss evolution of one-point PDF (Juszkiewicz et al. 1995,
Bernardeau \& Kofman 1995) or statistics of isodensity
contour, such as, genus or area statistics (Matsubara 1994). 
It is known that the skewness parameter $S_3$ obtained from numerical
simulations shows a  good agreement with  second-order prediction
up to regime
$\sigma^2\simeq 1$ ({\it e.g.} Hivon et al. 1995, Juszkiewicz et
al. 1995, Baugh,
Gazta\~{n}aga \& Efstathou 1995, {\L}okas et al. 1995). This fact is often used as a basis for
reliability of  second-order analysis.  

The skewness parameter has been also studied observationalily using
various galaxy surveys (see Table 1 of Hui \& Gazta\~{n}aga 1998).
As we cannot observe the dark matter distribution $\delta(\vex)$ directly,
the relation between the number density fluctuation of galaxies
$\delta_g(\vex)$ and that of underling matter
 $\delta(\vex)$ is crucially important to interpret
the observed data.   Their  relation is called  biasing and detailed theory
of galaxy formation is required to understand it quantitatively. If the
biasing relation is phenomenologically expressed as
\beq
\delta_g(\vex)=b_1 \delta(\vex)+\frac{b_2}2
(\delta(\vex)^2-\lla\delta(\vex)^2 \rra)+\cdots,
\eeq
with numerical coefficients $b_i$,  the first-order contribution of
skewness parameter for the galaxy distribution
 $\delta_g(\vex)$ becomes (Fry \& Gazta\~{n}aga
1994)
\beq
\frac{\lla \delta_g^3\rra}{\lla \delta_g^2\rra^2}=\frac1b_1 \lmk
S_3+\frac{3b_2}{b_1} \rmk. 
\eeq
When we discuss velocity dispersion as a function of galaxy overdensity
$\delta_g$, the coefficient $C$ becomes (Seto 2000a)
\beq
\frac{\lla \delta_g\veV^2\rra}{\lla \delta_g^2\rra \lla \veV^2\rra}=\frac{C}{b_1}.
\eeq
Note that the above expression does not depend on the coefficient $b_2$.

In the following subsections we study various second-order effects of
the moments $S_3$ and $C$ caused by smoothing operation.
We denote results derived in this subsection 
by $S_{3V}$ and $C_V$ with subscript $V$ to specify the 
smoothing method which we employ, i.e., 
the simple volume-weighted smoothing method. 

In Tables 1, 2 we present numerical
values  of  $C_V$ and $S_{3V}$ for power-law models smoothed by Gaussian
filter. 
   The parameter  $S_{3V}$ are given explicitly in
terms of the Hypergeometric functions (Matsubara 1994, {\L}okas et
al. 1995).  For the top-hat filter we obtain simple expression as
$S_{3V}= 34/7-(n+3)$ ({\it e.g.} Bernardeau 1994).

\subsection{Second-Order Correction for the  Mass-Weighted Smoothed Velocity
Field}

In the previous subsection we have discussed the smoothed (bulk) velocity
dispersion as a function of the smoothed density
contrast $\delta$. Following
 standard procedure of the Eulerian perturbation theory, we have adopted the 
volume-weighted smoothing method (eq.[\ref{vw}]). 
However it is not a simple
task to 
obtain the 
volume-weighted velocity field form N-body simulations due to their
Lagrangian nature (Bernardeau \& van de Weygart 1996, Kudlicki et
al. 2000).   In numerical analyses, a mass-weighted
smoothing is a
straightforward and convenient approach.  In the
followings we perturbatively 
investigate this approach  (Seto
2000b for details).

We
consider particles' system mimicing N-body simulations. Particles are
 initially placed at
grid points of  the orthogonal (Lagrangian) coordinate system $\veq$.
  We assume that each particle has equal mass $m(N)$ ($N$;
number of particles in a simulation box).   The simplest
numerical method for calculating  a  smoothed
velocity field (say at a point $\vex_0$) contains two steps.  The first
step is to sum up peculiar velocities of  particles 
$\veV(\vex(\veq_i))$ with   
weight $W(\vex(\veq_i)-\vex_0;R)$ determined by their
distances to the point $\vex_0$ in interest. 
The sum  represents  
``momentum density'' rather than  ``velocity''.
Thus the second step is to
 divide this  sum   with smoothed density at point 
$\vex_0$.
We denote the final 
 smoothed velocity field obtained in this manner by
$\veV_{mass}(\vex_0,R)$. It is expressed as follows:
\beq
\veV_{mass}(\vex_0,R)=\frac{m(N)\sum_i^N \veV(\vex(\veq_i))
W(\vex(\veq_i)-\vex_0;R)}{m(N)\sum_i^N W(\vex(\veq_i)-\vex_0;R)}.\lab{massvel}
\eeq
If we  increase the total number of particles $N$ with keeping their mean
density   
$\bar{\rho}$, we
can replace the summation $m(N)\sum_i^N$ by  three-dimensional
integral with respect to the 
 Lagrangian coordinate $\veq$. Next we change the coordinate system  from 
Lagrangian $\veq$ to  Eulerian $\vex$ and obtain the following result
as 
\beqa
\lim_{N\to\infty}m(N)\sum_i^N \veV(\vex(\veq_i))
W(\vex(\veq_i)-\vex_0,R)&=&\int d^3q \bar{\rho}W(\vex(\veq)-\vex_0;R)\veV(\vex(\veq))\\
&=&\int d^3 x W(\vex-\vex_0;R)\bar{\rho}\{1+\delta(\vex)\}\veV(\vex)\lab{mom}\\
&=&\bar{\rho}\{[\veV(\vex_0)]_R+[\veV(\vex_0)\delta(\vex_0)]_R\} ,\lab{mom2}
\eeqa
where we have used the well-known fact that   Jacobian $|\p \veq/\p
\vex|$ of the 
coordinate  transformation
$\veq\to\vex$ is given by the local density $1+\delta(\vex)$.
As shown in the factor $1+\delta(\vex)$ in equation (\ref{mom}), this
sampling is
mass-weighted. 
In the same manner the denominator of equation (\ref{massvel}) is given by
\beq
\lim_{N\to\infty}m(N)\sum_i^N
W(\vex(\veq_i)-\vex_0,R)=\bar{\rho}(1+[\delta(\vex_0)]_R).
\lab{den}\eeq
In this case we can obtain the volume weighted quantity.
With equations (\ref{mom}) and (\ref{den}) the smoothed velocity field is expanded
perturbatively as 
\beq 
\veV_{mass}(\vex,R)=[\veV_1(\vex)]_R+[\veV_2(\vex)]_R
+[\veV_1(\vex)\delta_{1}(\vex)]_R-[\veV_1(\vex)]_R[\delta_1(\vex)]_R+O(\sigma^3).
\eeq
The linear term $[\veV_1(\vex)]_R$ is unaffected by the  smoothing method.
The second term represents the  volume-weighted smoothed
velocity field at  the second-order of perturbations.
Let us extract out the new  second-order correction terms $\veV_{2m}(\vex,R)$ caused
by the mass weighted 
smoothing as 
\beq
\veV_{2m}(\vex,R)\equiv[\veV_1(\vex)\delta_1(\vex)]_R-[\veV_1(\vex)]_R[\delta_1(\vex)]_R.
\eeq
Thus  the velocity-density  moment $\lla \veV\cdot\veV\delta\rra$
for smoothed fields $\veV_{mass}$ and $\delta_R$ is written as 
\beq
\lla \veV_{mass}^2\delta_R\rra= \lla \veV_R^2\delta_R\rra +2  \lla \veV_{1R}\cdot
\veV_{2m}\delta_{1R}\rra .
\eeq
The first term is given in the previous subsection
 up to  the required order of $\sigma$ (see eq.[\ref{vd}]).
The additional term caused  by the mass-weighted smoothing  is evaluated
in terms of the power spectrum as 
\beqa
{2\lla [\veV]_R\cdot \veV_{2m}\delta_R\rra}&=&2H^2 f^2\int \frac{d^3k d^3l}{(2\pi)^6}
P(k)P(l) \lnk \frac1{k^2}+\frac{\vek\cdot\vel}{k^2l^2}\rnk\nn\\
& &\times  \lnk \exp\lmk-(k^2+l^2+\vek\cdot\vel) \rmk
- \exp\lmk-(k^2+l^2) \rmk\rnk.
\eeqa
For power-law models ($P(k)\propto k^n$) the correction term $\Delta C_M$
caused by the  above expression for the
factor $C$ is written with  Hypergeometric functions (Seto 2000b)
\beqa
\Delta C_M=\frac{2\lla [\veV]_R\cdot
\veV_{2m}\delta_R\rra}{\lla\veV_R^2\rra\lla\delta_R^2\rra}&=&
2F\lmk\frac{n+1}2,\frac{n+3}2,\frac32,\frac{1}4\rmk 
-\frac{(1+n)}3 F\lmk\frac{n+3}2,\frac{n+3}2,\frac52,\frac{1}4\rmk\nn\\
& &-2F\lmk\frac{n+1}2,\frac{n+3}2,\frac32,0\rmk.
\eeqa
The values of this correction term for various $n$'s are shown in 
table 1.

\subsection{Adaptive Smoothing}
It has been pointed out that  cosmic fields might be resolved more
efficiently by using adaptive smoothing filters than  traditional
spatially fixed filters ({\it e.g.} Springel et al. 1998). Seto (2000b,c)
studied second-order effects  of this method for
various moments of density and velocity fields. In the adaptive
method the  smoothing radius is determined by  local density. We use a
larger smoothing radius at the sparse underdense region but use a
smaller one  at  the high density region where nonlinear effects
would be larger.   Therefore,   the perturbative  treatment for
the adaptive method  might break down faster and its performance  should
be checked numerically.  In this subsection we
summarize second-order effects caused by  the adaptive smoothing 
 (Seto 2000b,c). Our
analytical results are compared with numerical ones in \S 4.

We first  determine the 
 adaptive smoothing radius so that number of particles within the
radius becomes constant. This condition is written as
\beq
R(\vex)^3[1+\delta(\vex)]_{R(\vex)}=R^3=const,
\eeq
where $R$ is the standard smoothing scale.
We can perturbatively solve this equation for the adaptive radius
 $R(\vex)$ and obtain
\footnote{In numerical analysis we use $R(\vex)=R(1+[\delta(\vex)]_R)^{-1/3}$.}
\beq
R(\vex)=R\lmk 1-\frac{[\delta(\vex)]_{R}}3+O(\sigma^2)   \rmk .
\eeq
There are mainly two approaches for the adaptive smoothing, namely, the
gather approach and the scatter approach (Hernquist \& Katz
1989).
Here we adopt the gather approach and  basically calculate smoothed
fields 
using operator $[\cdot]_{R(\vex)}$.  After some algebra we 
obtain 
the second-order correction for the smoothed  density field caused by
the  adaptive filter
\beq
\delta_{2A}(\vex,R)=-\frac{R}3 \delta_{1R}(\vex)\frac{\p}{\p R}
\delta_{1R}(\vex)+\frac{R}6\frac{d}{d R}\lla \delta_{1R}^2\rra,
\eeq
where the second term of the right-hand-side comes from  the requirement
$\lla \delta_{2A}\rra=0 $. 
This correction term gives new contribution $\Delta S_{3A}$ to the skewness 
parameter $S_3$ as
\beqa
\Delta S_{3A}&=&\frac{3\lla \delta_{2A}(\vex,R)\delta_{1R}(\vex)^2\rra }{\lla
\delta_{1R}^2\rra^2}\\
&=&-\frac{\lla
\delta_{1R}^2\rra R\p_R\lla\delta_{1R}^2\rra }{\lla\delta_{1R}^2\rra^2
}\\
&=&n+3,
\eeqa
where we have used relation $\lla \delta_{1R}^2\rra\propto R^{-(n+3)}$
for power-law models.
As we can see from above derivation, the term $\Delta S_{3A}$ does not
depend on 
the filter functions.

 In the same manner the additional
second-order terms for the smoothed velocity field is given as
\beq
\veV_{2A}(\vex,R)=-\frac{R^2}3 \delta_{1R}(\vex)\nabla^2 \veV_{1R}(\vex).
\eeq
We evaluate the new contribution $\Delta C_A$ for the coefficient $C$
caused by this correction term\footnote{Note that the contribution
$\lla\delta_{2A}\veV_1^2\rra$ vanishes.} and obtain following result for power-law
models as
\beqa
\Delta C_A&=&-\frac23 R^2 \frac{\lla \delta_{1R}^2 \rra\lla
\nabla^2\veV_{1R}\cdot \veV_{1R} \rra}{\lla \delta_{1R}^2 \rra\lla
\veV_{1R}^2 \rra}\\
&=&-\frac13\frac{\p \ln \lla \veV_{1R}^2 \rra}{\p \ln R}\\
&=&\frac{n+1}3,
\eeqa
where we have used relation valid for the Gaussian filter
\beq
\nabla^2\veV_{1R}(\vex)=-\frac1R \frac{\p  \veV_{1R}(\vex)  }{\p R},
\eeq
and a simple relation $\lla \veV_{1R}^2 \rra\propto R^{-(n+1)}$.

Since two second-order correction terms $\veV_{2m}$ and $\veV_{2A}$ are not
coupled, their effects on moments  $C$ are simply additive. For example, if we perform the  mass-weighted smoothing using the  adaptive
filter, the moment $C$ becomes
\beq
C=C_V+\Delta C_{M}+\Delta C_{A}.
\eeq
Same kind of additive relation holds for the skewness parameter.
\if0%%%%%%%%%%%%%%
where we have denoted the correction term caused by the adaptive smoothing  as
$\Delta C_A$. The subscript $A$ represents ``adaptive''.  In the same
manner we define the correction term $\Delta S_{3A}$ for the skewness
parameter $S$.

For power-law models $P(k)\propto k^n$ the correction terms  are given 
explicitly  as follows:
\beq
\Delta S_{A}=n+3,
\eeq
and
\beq
\Delta C_{A}=\frac{n+1}3.
\eeq
Numerical values for various $n$'s are shown  in tables 1 and 2.
\fi%%%%%%%%%%%%%%%%%

\section{SAMPLE VARIANCE}

In the next section we numerically evaluate the factors $C$ or $S_3$ using
N-body simulations. For each run of simulations we calculate the volume
average $A(Y,\calv)$ of  a field $Y(\vex)$ within the simulation box
(cube) $\calv$ as follows:
\beq
A(Y,\calv)\equiv\frac1\calv \int Y(\vex) d^3x.\lab{ves}
\eeq
The ensemble average $\lla A(Y,\calv) \rra $ of the 
estimator $A(Y,\calv)$ coincides with
the ensemble average  $\lla Y(\vex)\rra$ of the original field $Y(\vex)$ and we
have 
\beq
\lla A(Y,\calv) \rra=\lla Y(\vex)\rra.
\eeq
But the measured value $A(Y,\calv)$ fluctuates around its mean value 
$\lla Y(\vex)\rra$ due to  finiteness of the simulation box. This
fluctuation can become important for analysis of  numerical simulations. We
call the  root-man-square value $D(Y,\calv)$ of this fluctuation
 as the sample variance 
 \beq
D(Y,\calv)^2\equiv \lla (A(Y,\calv)-\lla Y\rra)^2 \rra.
\eeq
In this subsection we discuss the sample variances of various moments
following Seto \& Yokoyama (2000).
The sample variance is closely related to the number of statistically
independent 
regions that is roughly determined by the correlation length of the field
$Y(\vex)$  and the size of the simulation box (or survey volume).
Effects of the sample variance is generally more important for the
velocity field than the density field, as the former is more weighted to 
large-scale fluctuations and has a  larger correlation length.

For fields
$Y(\vex)=\veV(\vex)^2\delta(\vex)$ and 
$Y(\vex)=\delta(\vex)^3$   we have vanishing means $\lla\delta(\vex)
\veV(\vex)^2\rra=0$ and $\lla \delta(\vex)^3\rra=0$ at the linear
order of perturbations as explained before\footnote{In this section we drop the subscript $R$
 (smoothing radius) and simply denote
$\veV(\vex)=[\veV(\vex)]_R$ and $\delta(\vex)=[\delta(\vex)]_R$. }.
However their sample variances
do not vanish at the linear  order and we have the following relation
for relative fluctuations of their measured value $A(Y,\calv)$
\beq
\frac{D(Y,\calv)}{\lla Y \rra }=O(\sigma^{-1}). \label{sam1}
\eeq
In contrast both
 the expectation values and the sample variances  for the second moments
$Y(\vex)=\delta(\vex)^2$ or $Y(\vex)=\veV(\vex)^2$   have contributions
from 
the linear modes
and we have
\beq
\frac{D(Y,\calv)}{\lla Y \rra }=O(1). \label{sam0}
\eeq 
  In the weakly
nonlinear regime, 
 the sample variance of the factor $C\equiv \lla
\veV^2\delta\rra/(\lla\veV^2\rra \lla\delta^2\rra)$ 
  would be
dominated by  fluctuations of the numerators  $\lla \veV^2
\delta\rra$, as  expected  from equations (\ref{sam1}) and (\ref{sam0}).

Here we evaluate the sample variance $D(Y,\calv)$  caused by the linear
modes  for the moment
$Y(\vex)=\veV(\vex)^2\delta(\vex)$. We assume that
our simulation box is a cube with side length $L$.
After some algebra  the sample variance can be written as 
\beqa
D(\veV^2 \delta,\calv)^2
&=&\frac{2 H^4}{L^{9}} \sum_\vek \sum_\vel P(k)P(l)P(|\vek+\vel|)
W(kR)^2W(lR)^2W(|\vek+\vel|)^2 \nn\\
& &\times
\frac{\vek\cdot\vel}{k^2l^2}\lnk\frac{\vek\cdot\vel}{k^2l^2}
-2\frac{\vek\cdot(\vek+\vel)}{k^2(\vek+\vel)^2}   \rnk, 
\eeqa
where the wave vector $\vek$ is expressed as $\vek=2\pi/L(n_x,n_y,n_z)$
with integers $n_x$, $n_y$ and $n_z$.
Now let us approximate the above
expression by replacing the summations with integrals, because it is not
easy to evaluate these summations.
 We introduce the large-scale cut-off at wave-number
$k_{min}= 2\pi/L$ that is determined by the side length $L$ of the
simulation box $\calv$. Then we obtain
\beqa
D(\veV^2 \delta,\calv)^2 
&\simeq &\frac{2H^4}{(2\pi)^6L^3}\int_{k_{min}}d^3 k\int_{k_{min}}d^3 l 
  P(k)P(l)P(|\vek+\vel|)
W(kR)^2W(lR)^2W(|\vek+\vel|)^2\nn\\
& &\times\frac{\vek\cdot\vel}{k^2l^2}
\lnk\frac{\vek\cdot\vel}{k^2l^2}
-2\frac{\vek\cdot(\vek+\vel)}{k^2(\vek+\vel)^2}   \rnk .
\eeqa
Due to the rotational symmetry around the origin
this integral is simplified to  a three-dimensional integral. In the
same manner the sample variance for the third-order moments of density
field  is given as
follows: 
\beq
D(\delta^3,\calv)^2 
\simeq \frac{6}{(2\pi)^6L^3}\int_{k_{min}}d^3 k\int_{k_{min}}d^3 l 
  P(k)P(l)P(|\vek+\vel|)
W(kR)^2W(lR)^2W(|\vek+\vel|)^2.
\eeq
For the second moments $\lla\veV(\vex)^2\rra$ or $\lla\delta(\vex)^2\rra$ we have the 
following results
\beq
D(\veV^2,\calv)^2\simeq \frac{2H^4}{(2\pi)^3L^3}\int_{k_{min}} d^3k P(k)^2W(kR)^4k^{-4},~~~
D(\delta^2,\calv)^2\simeq \frac{2}{(2\pi)^3L^3}\int_{k_{min}} d^3k P(k)^2W(kR)^4.
\eeq

In this section we made a crude estimation of the sample variance
$D(Y,\calv)$.  We have only studied the fluctuations
caused by the linear modes and used approximation of replacing
summations with integrals.  Therefore our estimation is not quantitatively
correct. However the present analysis would be useful to interpret numerical
results given in the next section.

\section{NUMERICAL ANALYSIS}
We performed P${}^3$M N-body simulations using  $64^3$ particles in a
cube with 
side length $L=1$. 
The numerical code is provided by Couchman.
We firstly  investigated two scale-free models
with 
spectral indexes
$n=1$ and $0$ in  Einstein
de-Sitter background and  run several
simulations for each models.   We stopped calculations at
the epoch 
when the  matter fluctuation at the smoothing radius
 $R=1/80$ ($R$: the smoothing radius for the
Gaussian filter)  becomes $\sigma_R^2\simeq 2$ for the $n=0$ model and
$\sigma_R^2\simeq 1$  for the $n=1$ model.
 Before describing numerical results, we mention two important effects
that must be cared here.  The first one is the sample variance described
in the previous section and the second one is the artificial cut-offs
induced in the initial fluctuations.

We are interested in the velocity and density fields at weakly nonlinear
regimes. As explained before, the sample variances of various statistical
quantities become larger for a larger smoothing radius. In figure 1 we
estimated the relative fluctuations $D(Y,\calv)/\lla Y\rra$ caused by
the linear modes using expressions given in \S 3 at the final epoch of
each simulation. 
Apparently, the moment $\lla \veV^2\delta\rra$ has the  largest
fluctuations. The fluctuation can be as much as the expectation value
itself, i.e., $100\%$ variance for smaller  $\sigma^2(R)$.
The sample variance is very
sensitive to the large-scale fluctuations. Roughly speaking, the sample
variance  is 
more important 
 for $n=0$ model than for $n=1$ model, as the former has more
large-scale powers.  Because of the same reason, the velocity dispersion
$\lla \veV^2\rra $ has
larger fluctuations than the variance of density fluctuations 
$\lla \delta^2\rra$.

There are two cut-off scales for initial matter fluctuations in our
simulations. One is the Nyquist frequency $k_{Nyq}=\pi N/L$ determined by
the initial separation of particles and the other one is the wave number
$k_{min}$ determined by the side length of the simulation box as $k_{min}=2\pi
/L $. Our analytical predictions given in \S 2  might be
different from numerical results due to these  artificial cut-offs.
 To estimate their effects  we calculate the   factors $C_V$ and $\Delta C_M$
with a modified power spectra $P_{eff}(k)$ mimicing N-body simulations as
follows ({\it e.g.} Seto 1999): 
%%%%%%%%%%%%%%%%%%%%%%%%%%%%%%%
\beqa P_{eff}(k)=\cases{
0 & $(0\le k<k_{min})$ \cr
k^n &  $(k_{min}\le k\le k_{Nyq} )$ \cr
0  &  $(k_{Nyq}< k  )$. \cr
}\eeqa
%%%%%%%%%%%%%%%%%%%%%%%%%%%%%%
Analysis for the $n=-1$ model  is particularly
interesting as we have $C_V=\Delta C_M=\Delta C_A=0$ and velocity dispersion does not depend
on the density contrast even at the second-order of perturbation. However,
the wave-number 
 integral (eq.[\ref{v2}])
 for the velocity dispersion diverges in the $k\to 0$ limit and the
velocity field would be  strongly  affected by the artificial large-scale
 cut-off  at 
$k_{min}$. Actually we have confirmed numerically  that the factors $C_V$
and $\Delta C_M$ for this model
with the modified spectrum $P_{eff}(k)$ would
 considerably deviate from the original scale-free  model.
We apply  this analysis for models with indexes
 $n=0$ and $ 1$, and found that
the factors $C_V$ and $\Delta C_M$ could deviate largely from 
predictions for the original scale-free models at  $R\lsim 1/80$ or
$R\gsim 1/20$. Thus we limit our analysis
 only for smoothing radius  with $1/80\le R\le 1/20$.

We have  analyzed N-body data at the final  epoch for each run. 
As our
simulations are basically self similar, we can expect that physics is
characterized only by the magnitude of nonlinearity $\sigma^2(R)=\lla
\delta_R^2 \rra$. Thus
we use the variance $\sigma^2(R)$ rather than the radius
 $R$ to represent the scale dependence. 
 For $n=0$ model we have performed three runs. Our numerical results are
given in figure 2.  In the upper left panel the factors $S_{3V}$ and $C$ are
shown as a function of nonlinearity $\sigma^2(R)$. The symbols represent
the average values of three runs and the error bars are the variances of
their 
fluctuations.   The magnitude of error bars is roughly understood
by checking the sample variances caused by the linear modes given in
figure 1 where we have added numerical results from simulations for the
quantity 
$\lla \veV^2\delta\rra$.

 As shown in literatures ({\it e.g.} Hivon et al. 1995, Juszkiewicz et
al. 1995, {\L}okas et al. 1995), the skewness parameter $S_{3V}$ at
$\sigma^2(R)\lsim 1$  is nearly constant for $n=1$ model and close to the predictions of
the second-order perturbation theory $S_{3V}=3.14$. In contrast the factor
$C$ depends strongly on nonlinearity $\sigma^2(R)$ and we have $C\simeq
0.05$ at $\sigma^2(R)\simeq 1$. This value is only $\simeq 40 \%$ of the
second-order prediction $C=0.12$.  The numerical results for $C$ become
close to the analytical prediction at $\sigma^2(R)=0.2$ where the sample
variance is fairly large.  
 In figure 2 we also show  the 
constrained velocity dispersion
$\Sigma_V^2(\delta)/\sigma_V^2$ in
the range of density contrast  $-1.5\sigma(R)\le \delta \le
1.5\sigma(R)$. We have made a  same kind
of averaging operation for three runs as in the case of the factors $S_{3V}$ and $C$.
The dashed lines are the analytical predictions given in equation
(\ref{convel})
whose slope is $C\simeq 0.12$.  The numerical results is close to the
analytical prediction at small $\sigma^2(R)$, but the slope of numerical
results become smaller for larger  $\sigma(R)$ as expected form 
behavior of the factor $C$.

Let us show same  analyses for the model with $n=1$. 
For this model we performed
seven runs. Results are shown in
 figure 3.  The measured parameter $S_{3V}$ is nearly constant  but
somewhat ($\simeq 10\%$) smaller than the analytical prediction
$S_{3V}=3.03$.  The 
factor $C$ is again  a decreasing function of nonlinearity
$\sigma^2(R)$.  However its convergence to the analytical prediction $C=0.2$ 
 seems to be slower. Even at $\sigma^2(R)=0.04$ where the sample variance
is considerably large, the measured value is much smaller than the
analytical one. The constrained velocity dispersion $\Sigma^2_V(\delta)$ 
behaves in the same manner as the model with $n=0$. The effective slope becomes
smaller for larger $\sigma^2(R)$.
The difference between the second-order prediction and the numerical
results is caused by the nonlinear effects that cannot be traced by the
second-order perturbation. For
given $\sigma^2(R)$ this difference is larger for the model with $n=1$
than the one with $n=0$. 
This is reasonable as the former has  more small scale power
initially and nonlinear effects would be larger.  Nagamine, Ostriker \&
Cen (2000) have numerically studied the cosmic Mach number (Ostriker \&
Suto 1990) as a function
of 
overdensity using an LCDM hydrodynamical simulation. They have provided
two-dimensional contour plots for the probability distribution functions
of the smoothed  dark-matter overdensity $\delta_R$
and the bulk velocity $|\veV_R|$ at various smoothing scales $R=5, 10$ and
$20\hmpc$ (see their figures 10-12). Comparing these figures we can
observe the scale dependence of $\Sigma_V^2(\delta)$ discussed so far.

We have also performed N-body simulations (5 runs) for a  CDM model.
Our model parameters are fixed at $\Omega_0=1.0$, $h=0.5$  \footnote{As commented in \S 2.1,  our results for the
parameter $C$ would
depends weakly on the background  cosmological parameters.} and the CDM
shape parameter 
$\Gamma=0.5$. We used the  CDM transfer
function of  Bardeen et al. (1986) with the primordially $n=1$ spectrum
and normalized the linear spectrum by $\sigma_8=0.8$ ($\sigma_8$ is the
rms density fluctuation within a top-hat sphere of radius 8$h^{-1}$Mpc).
As the velocity field is heavily  weighted to large-scale fluctuations,
we 
have adopted  a $L=400 h^{-1}$Mpc simulation box.
The prediction $C$ by the second-order theory at the weakly nonlinear regime
is 
$C\sim 0.13$ and close to   the previous  result for the $n=0$ model.
Our numerical results are  $C=0.094\pm0.06$ for $R=10h^{-1}$Mpc
($\sigma^2(R)=0.07$),  and  $C=0.068\pm0.01$ for $R=5h^{-1}$Mpc
($\sigma^2(R)=0.37$), which are 
consistent with the prediction of the second-order theory.   
The mean values and the error bars depend on 
nonlinearity $\sigma_R^2$ as in the case of the $n=0$ model.

Hui \& Gazta\~{n}aga (1999) have pointed out that a nonlinear combination
$f$ of estimators $A(Y_i,\calv)$ (see eq.[\ref{ves}]) should be analyzed with sufficient care. 
We have following inequality even in the case of the unbiased estimators
$\lla 
A(Y_i,\calv)\rra=\lla Y_i \rra$ :
\beq
f(\lla Y_1\rra,\cdots,\lla Y_n\rra)=f(\lla A(Y_1,\calv)\rra,\cdots,\lla
A(Y_n,\calv)\rra)\neq\lla f( A(Y_1,\calv),\cdots,
A(Y_n,\calv)) \rra. \lab{nld}
\eeq
This inequality leads to a biased estimation for the nonlinear
function $f(\lla Y_i \rra)$. 
When the variance $\bar{\xi}^\calv_2$ of density contrast smoothed in
survey 
volume $\calv$ is much smaller than unity, this bias $\Delta S_{3V}$ for the estimation of the
skewness $S_{3V}$ is approximately given as follows (Hui \&
Gazta\~{n}aga 1999) 
\beq
\frac{\Delta S_{3V}}{S_{3V}}=\alpha_1
\frac{\bar{\xi}^\calv_2}{\lla \delta_R^2\rra}+\alpha_2 \bar{\xi}^\calv_2
 ,\lab{51}
\eeq
where the coefficients $\{\alpha_i\}$ are determined by the power
spectrum. 
 As we use data in full simulation boxes, there are no mean density
fluctuations $\bar{\xi}^\calv_2=0$ and the above expansion (\ref{51})
vanishes. Therefore we evaluate the estimation bias by 
directly comparing numerical values $f(\lla A(Y_1,\calv)\rra,\cdots,\lla
A(Y_n,\calv)\rra)$ and $\lla f( A(Y_1,\calv),\cdots,
A(Y_n,\calv)) \rra$ (the ensemble average $\lla\cdot\rra$ is taken for
different realizations of simulations). We find that differences between
these two are very small both for $S_V$ and $C_V$. For data presented in
figures 2 and 3 the difference is smaller than $1\%$. This is due to the
fact that the fluctuations of their denominators (expressed by $\lla
\delta^2\rra$ and $\lla\veV^2\rra$) are very small as expected
from figure 1.     This estimation bias would be more important in
actual observational situations.

In figure 4 we present numerical results for the factors $S_3$ and $C$
that were evaluated using the adaptive filter described in \S 2.3.  As
shown in the left panels, the skewness parameter $S_3$ shows reasonable
agreement with the second-order perturbation theory. But the parameter
$C$ again shows 
strong 
dependence on nonlinearity $\sigma^2(R)$ even in this regime and its
convergence to the analytical prediction is slow. These behaviors of
$S$ and $C$ are similar to the previous 
case of non-adaptive smoothing.    The
parameter $C$ for 
the model with $n=0$ becomes close to the second-order result
$C=0.45$. 
For $n=1$ model,
the second-order correction $\Delta C_A=0.67$ caused by the
adaptive smoothing seems not to  match the numerical result even at the
scale 
$\sigma^2\simeq0.04$.  
 In the adaptive  method we use
a smaller smoothing radius at high density regions.  Thus numerical
results are expected to
 be more contaminated by strongly nonlinear effects than
the previous simple smoothing method especially for $n=1$ model.
The numerical result is estimated as $\Delta
C_A\simeq 0.44$ from $C_V+\Delta C_M+\Delta C_A=0.56$ (figure 4) and $C_V+\Delta C_M=0.12$  (figure 3).

Finally we comment on possibility of the determination 
of the parameter $C_V$ by the actual  observations.  As only 
 the line-of-sight velocity field $V_{||}(\vex)$ is  observable, 
we evaluate the fluctuations  of linear modes  for the  moment  $\lla
V_{||}^2\delta\rra$.  For this evaluation we cannot simply 
 apply  results derived under  the periodic boundary condition in \S
3.  We estimate
the sample variance $D(V_{||}^2\delta,\calv)$ using  a method similar  to Seto \& Yokoyama
(2000).  As an example we have examined  a flat
cold-dark-matter model with cosmological parameters   $h=0.7$, 
$\Omega_0~=0.3$ and $\lambda_0=0.7$. Normalization of the power spectrum is
determined by abundance of rich clusters as $\sigma_8=0.50
\Omega_0^{-0.53+0.13\Omega_0}$ (Eke, Cole \& Frenk
1996). The Gaussian smoothing radius is fixed at
$R=12\hmpc$. The second-order
prediction of the  parameter $C$ for the primordially scale-invariant 
 CDM power spectrum  in
a weakly nonlinear regime is close to that for the $n=0$
model. Therefore we assume
\beq
\lla V_{||}^2\delta\rra
=\frac{\lla\veV^2\delta\rra}3\simeq \frac{C\lla
\veV^2\rra\lla\delta^2\rra}3 \simeq 0.04  \lla
\veV^2\rra\lla\delta^2\rra.
\eeq
Using these relations  we obtain $D(V_{||}^2\delta,\calv)/\lla
V_{||}^2\delta\rra\sim8$  for a spherical volume with
survey depth $80 \hmpc$. Thus we can hardly measure the
parameter $C$ from current catalogs of peculiar
velocity field, such as, Mark III (Willick et al. 1997)
or SFI (Haynes et al. 1998).
The relative fluctuation becomes close to unity only at a survey
volume with depth $\sim 300 \hmpc$. In the actual observational
situations the effects caused by the nonlinear estimators
(eq.[\ref{nld}]) would be also important.
\section{SUMMARY}

We have investigated various statistical quantities of smoothed velocity 
and density fields. Analytical predictions based on second-order
perturbation theory are compared with numerical results obtained from
N-body simulations. Our primary target is the velocity dispersion
$\Sigma^2_V(\delta)$ as a 
function of local density contrast $\delta$ at weakly nonlinear scales.
At the linear order with isotropic random Gaussian initial condition the
dispersion 
$\Sigma^2_V(\delta)$ does not depend on $\delta$. However second-order
modes generate a correction term proportional to $\delta$ with its slope
$C\equiv \lla \veV^2\delta\rra/(\lla \veV^2\rra \lla\delta^2\rra)$.
We have confirmed this dependence at small $\sigma(R)$ and found that the parameter $C$
depends largely on nonlinearity $\sigma(R)$ even in the regime where
the skewness parameter $S_3$ is nearly constant value and close to
prediction by second-order perturbation theory.

We have also studied  second-order effects caused by
various smoothing methods. Proper
treatment of smoothing is crucial to quantitatively 
discuss weakly nonlinear effects of
cosmic fields. 
A mass weighted smoothing is convenient to analyze velocity field traced
by particles as in N-body simulations. We have basically used this method
for our numerical analysis.
We found that the parameter $C$ actually
 approaches to the
value that includes effects of mass-weighted smoothing
in the limit of small $\sigma(R)$.  But a large smoothing
radius is required to recover the second-order
result. For example we have to take
the radius $R$ with $\sigma^2(R) \lsim 0.1$ for the $n=0$ model and
$\sigma^2(R) \lsim 0.01$ for the $n=1$ model.  Adaptive
smoothing methods might be useful to resolve cosmic field especially in
underdense regions.  The skewness parameter $S_3$ obtained numerically
with an adaptive method (gather approach) is closed to the corresponding
second-order prediction up to $\sigma^2(R)\simeq 1$.
The skewness parameter and its generalizations are closely related to
weakly nonlinear evolution of genus or area statistics of isodensity
contour (Matsubara 1994) for which the adaptive method would be effective
(Springel et al. 1998). Our result is encouraging for application of
second-order analysis for density field smoothed by the adaptive filter.

It has been known that the second-order predictions at $\sigma(R)\simeq
1$  are more accurate for the skewness of the real space density field
than that of the velocity divergence field $\propto \nabla\cdot \veV$
(Juszkiewicz et al. 1995)
or  the redshift space density field  (Hivon et al. 1995). But we should
notice  that calculation  of the coefficient $C\equiv \lla \veV^2\delta\rra/(\lla \veV^2\rra \lla\delta^2\rra)$ directly requires
information of the peculiar velocity field as  a vector field.  
 We found that performance of the second-order prediction is generally
 worse for
the factor $C$  
 than for the skewness
 $S_3$.     This fact seems interesting. As the density field
is more weighted to the  small-scale fluctuations than the velocity
field, we 
can naively expect that highly nonlinear effects would be stronger for
the density field. Therefore our numerical investigation provides 
applications of the second-order perturbation
theory with an important caveat.

\acknowledgements
We thank Naoki Yoshida for discussion about numerical schemes and  an
anonymous referee for valuable comments 
to improve this manuscript.
This work is  supported by Japanese  Grant-in-Aid for Science Research
Fund of the Ministry of Education, Science, Sports and Culture Grant
Nos. 0001461 and  11640235.

\newpage

\begin{center}
TABLE 1\\
{\sc velocity-density moment $C$ (Gaussian filter)}\\
\ \\
\begin{tabular}{ccccccc}
\hline\hline
 spectral index $n$  &-1& 0 & 1& 2 \\
\hline
 $C_V$  ~~ &~~ 0~~ &~~  0.24  ~~ &~~0.31 ~~ &~~0.38 \\
 $\Delta C_M$  ~~ &~~ 0~~ &~~ -0.12~~ &~~-0.11 ~~ &~~0.053\\
 $\Delta C_{A}$  ~~ &~~ 0~~ &~~ 0.33~~ &~~0.67 ~~ &~~1.0\\
 $C_V+\Delta C_M$  ~~ &~~ 0~~ &~~ 0.12~~ &~~0.20 ~~ &~~0.43\\
 $C_V+\Delta C_M+\Delta C_A$  ~~ &~~ 0~~ &~~ 0.45~~ &~~0.87 ~~ &~~1.43\\
\hline
\end{tabular}
\end{center}

\begin{center}
TABLE 2\\
{\sc skewness $S_3$ (Gaussian filter)}\\
\ \\
\begin{tabular}{ccccccc}
\hline\hline
 spectral index $n$  &-1& 0 & 1&2 \\
\hline
 $S_{3V}$  ~~ &~~ 3.5~~ &~~  3.1  ~~ &~~3.0~~&~~3.1  \\
 $\Delta S_{3A}$  ~~ &~~ 2.0~~ &~~ 3.0~~ &~~4.0~~ &~~5.0\\
 $S_{3V}+\Delta S_{3A}$  ~~ &~~ 5.5~~ &~~ 6.1~~ &~~7.01~~ &~~8.1 \\
\hline
\end{tabular}
\end{center}

\newpage
%\include{ref}
%\begin{references}

\begin{figure}
\epsscale{1}
\plotone{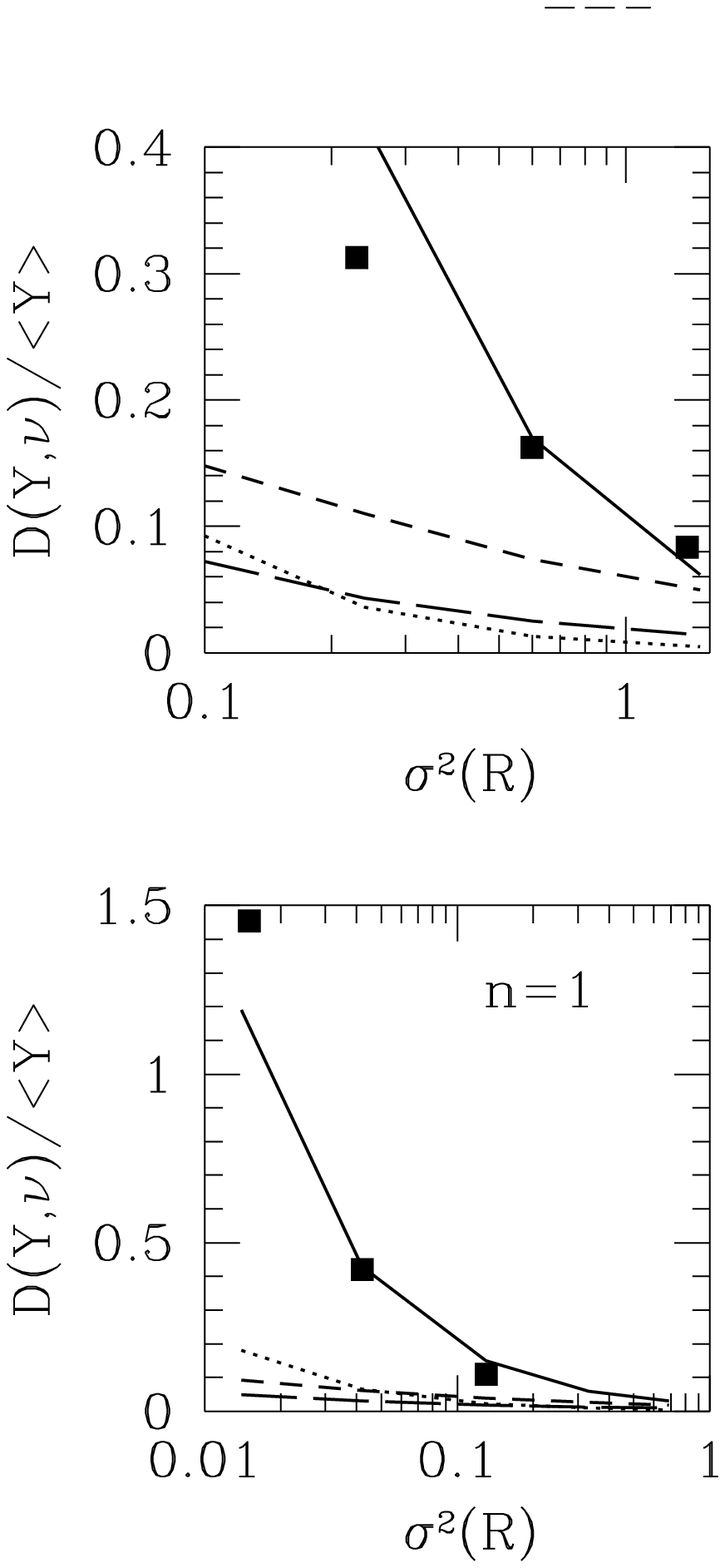}
\caption{Relative fluctuations  caused by the linear modes in our
 simulations (see \S 3 for their estimation). The solid lines correspond
	   to the field 
 $Y(\vex)=\veV(\vex)^2\delta(\vex)$, the dotted  to
 $Y(\vex)=\delta(\vex)^3$, the short-dashed to $Y(\vex)=\veV(\vex)^2$
 and the long-dashed to   $Y(\vex)=\delta(\vex)^2$. We numerically
 estimate the fluctuations $D(\veV^2\delta,\calv)$ from different
 realizations of simulation (see figures 2 and 3) and plot its relative
 magnitude
 with 
 respect to the expectation value $\lla\veV^2\delta\rra$  (predicted by the
 second-order perturbation theory) by the filled squares.} 
\end{figure}

\begin{figure}
\epsscale{1}
\plotone{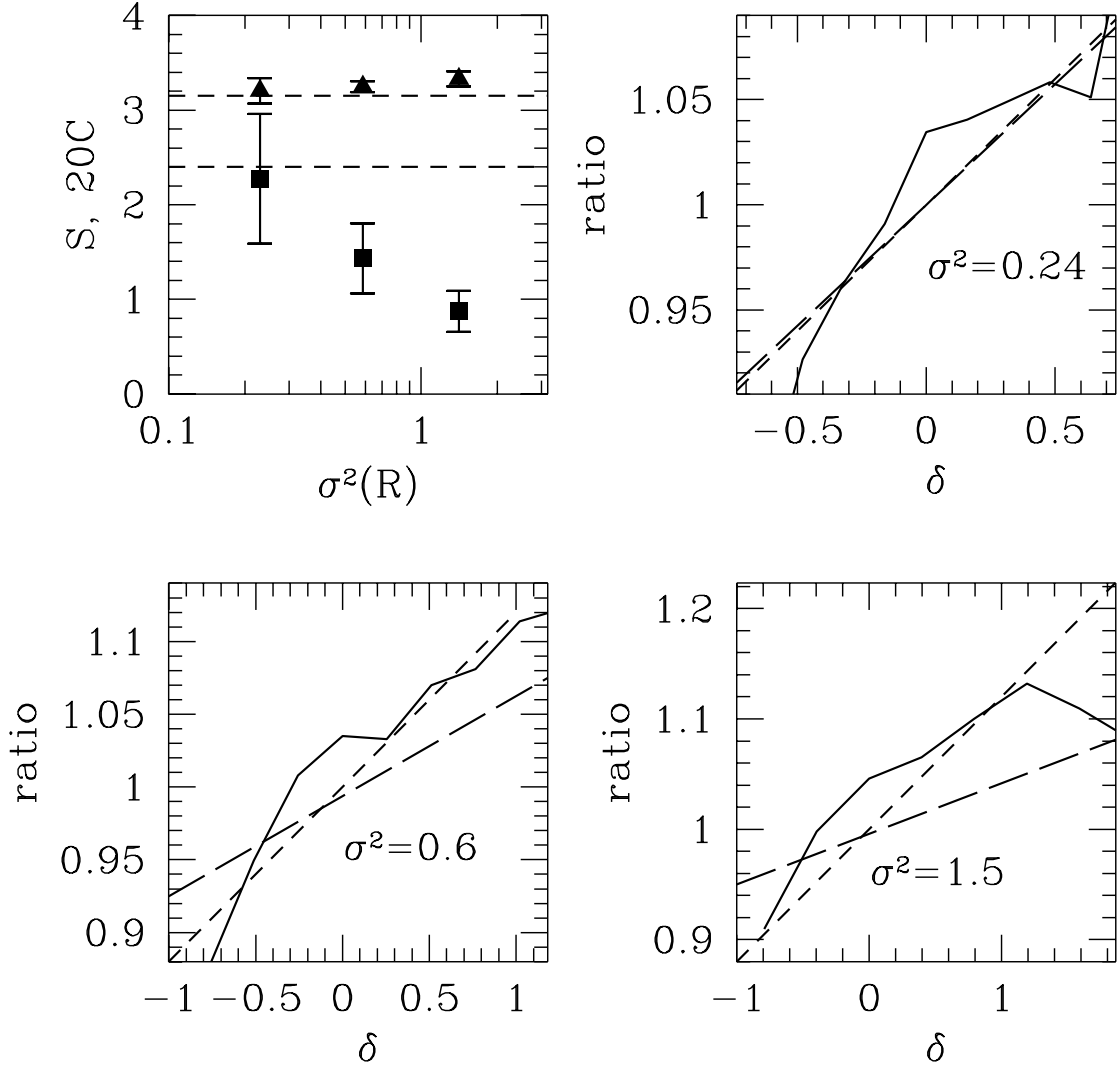}
\caption{Two factors $C$,  $S_{3V}$ and the constrained bulk velocity
	   dispersion $\Sigma^2_V(\delta)$ for
	   $n=0$ 
	   model. (1) Upper-left panel: The mean values (filled
	   symbols) and error bars are obtained from three runs of
	   simulations.  The squares represent the factor $C$ in the form
	   of $ 20C$, and the triangles the skewness $S_{3V}$. Predications by
	   the second-order perturbation theory are shown by the dashed
	   lines  ($C=0.12$ and
	   $S_{3V}=3.14$).  (2) Other   three panels: The constrained
	   velocity dispersion is presented using the ratio
	   $\Sigma^2_V(\delta)/\sigma_V^2$ in the range 
           $-1.5 \sigma(R)\le \delta \le 1.5 \sigma(R)$.  The short-dashed
	   lines  are    
	   predictions by the Edgeworth expansion method with
	   the factor $C$  from second-order perturbation theory, and
	   the long-dashed lines are drawn with $C$ obtained from
	   numerical simulations.}
\end{figure}

\begin{figure}
\epsscale{1}
\plotone{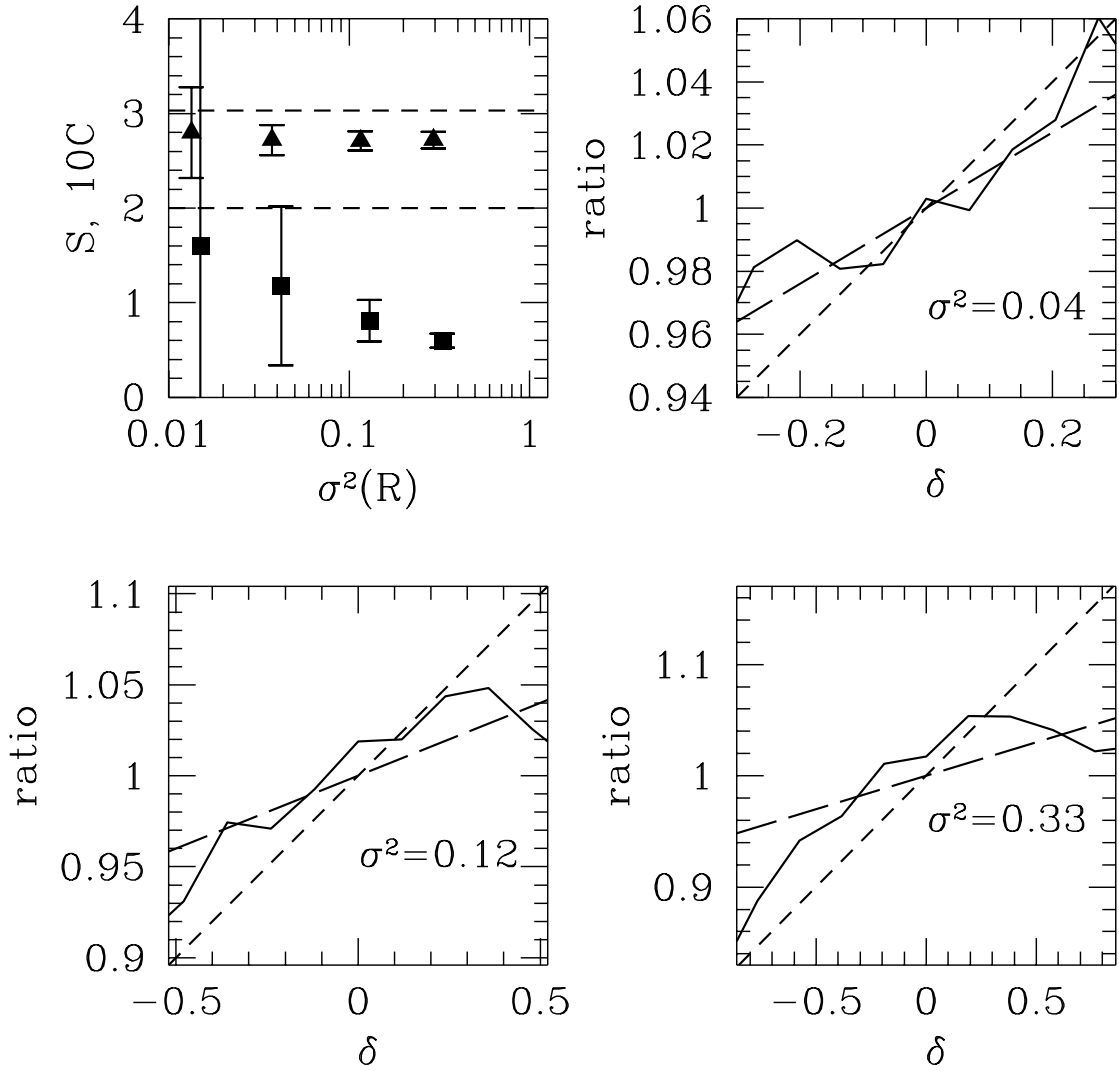}
\caption{Numerical results for $n=1$ model. Figures are given in the same manner
	   as figure 2.  Second-order perturbation predicts $C=0.20$ and
	   $S_{3V}=3.03$.}
\end{figure}

\begin{figure}
\epsscale{1}
\plotone{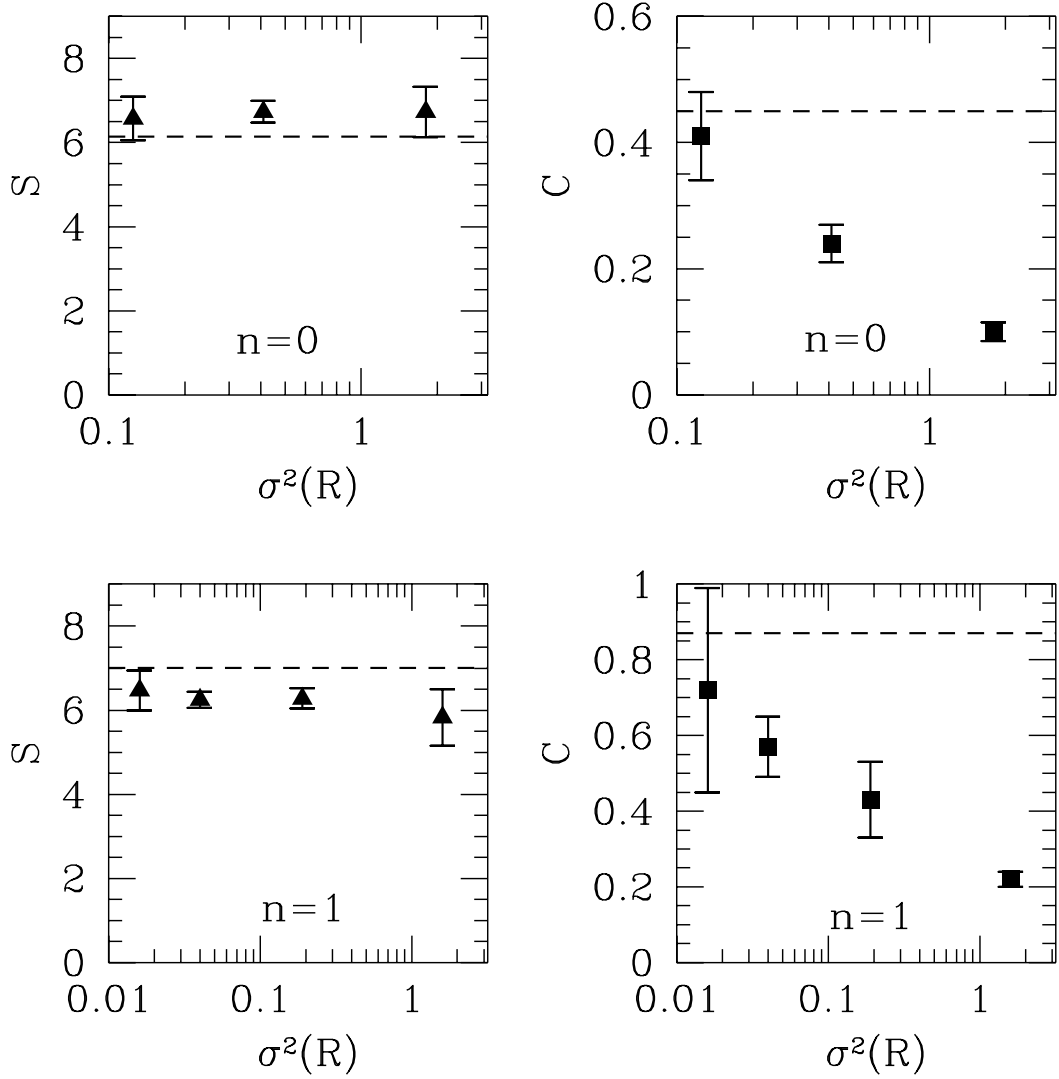}
\caption{Effects of the adaptive smoothing for moments $S_3$ and $C$. We
	   present the averaged values as in Figures 2 and 3.  The upper
	   panels correspond to $n=0$ model and lower panels to $n=1$
	   model.  The dashed lines represent the second-order
	   predictions: $S_3=6.1$, $C=0.45$ for $n=0$ model and $S_3=7.0$,
	   $C= 0.87$ for $n=1$ model.}
\end{figure}

\end{document}